\documentclass[10pt]{iopart}
\usepackage{iopams}
\usepackage{graphicx}
\usepackage{cite}
\newcommand{\text}[1]{\mathrm{#1}}
\def\ds{\displaystyle}
\usepackage{eurosym}
\begin{document}

\title[First 3D AWAKE simulation]{First fully kinetic three-dimensional simulation of the AWAKE baseline scenario}

\author{N. Moschuering}
\address{Ludwig-Maximilians-Universit\"{a}t, 80539 Munich, Germany}
\ead{Nils.Moschuering@physik.uni-muenchen.de}
\author{K.V. Lotov}
\address{Budker Institute of Nuclear Physics, Novosibirsk, 630090, Russia}
\address{Novosibirsk State University, Novosibirsk, 630090, Russia}
\ead{K.V.Lotov@inp.nsk.su}
\author{K. Bamberg}
\address{Ludwig-Maximilians-Universit\"{a}t, 80539 Munich, Germany}
\ead{Karl-Ulrich.Bamberg@physik.uni-muenchen.de}
\author{F. Deutschmann}
\address{Ludwig-Maximilians-Universit\"{a}t, 80539 Munich, Germany}
\ead{Fabian.Deutschmann@physik.uni-muenchen.de}
\author{H. Ruhl}
\address{Ludwig-Maximilians-Universit\"{a}t, 80539 Munich, Germany}
\ead{Hartmut.Ruhl@physik.uni-muenchen.de}

\vspace{10pt}
\begin{indented}
\item[]\today
\end{indented}

\begin{abstract}
The ''Advanced Proton Driven Plasma Wakefield Acceleration Experiment`` (AWAKE) aims to accelerate leptons via proton-beam-driven wakefield acceleration. It comprises extensive numerical studies as well as experiments at the CERN laboratory. The baseline scenario incorporates a plasma volume of approximately $62\,\text{cm}^3$. The plasma wavelength is about $1.25\,\text{mm}$ and needs to be adequately resolved, using a minimum of $130$ points per plasma wavelength, in order to accurately reproduce the physics. The baseline scenario incorporates the proton beam micro-bunching, the concurrent non-linear wakefield growth as well as the off-axis electron beam injection, trapping and acceleration. We present results for the first three-dimensional simulation of this baseline scenario with a full model, using a sufficient resolution. The simulation consumed about $22\,\text{Mch}$ of computer resources and scaled up to 32768 cores, thanks to a multitude of adaptions, improvements and optimization of the simulation code PSC. Through this large-scale simulation effort we were able to verify the results of reduced-model simulations as well as identify important novel effects during the electron injection process.
\end{abstract}

% Uncomment for PACS numbers
%\pacs{00.00, 20.00, 42.10}
\vspace{2pc}
\noindent{\it Keywords}: plasma wakefield acceleration, proton driver, numerical simulations

\ioptwocol
\section{Introduction}
Proton-beam-driven plasma wakefield acceleration promises to increase the lepton energy which is available to scientific laboratories \cite{PoP18-103101,RAST9-85,EPJC76-463}.
The first step towards future colliders based on this principle is the AWAKE experiment at CERN \cite{NIMA-829-3,NIMA-829-76,PPCF60-014046}, which has already demonstrated controlled self-modulation of the proton beam \cite{PRL122-054801,PRL122-054802} (the key element of the concept) as well as electron acceleration in the proton-driven wakefield \cite{Nat.561-363}.

In order to better understand the physics of proton and electron beam interaction with plasma, a massive simulation campaign accompanied the experiment from its very early stages \cite{JPP78-347, PRAB21-011301, PoP20-103111, PRL109-145005, PoP21-056705, PoP18-103101, PoP20-013102, PRST-AB16-041301, PoP20-083119, PRL112-194801, PPCF56-084013, PoP21-083107, PoP21-123116, NIMA-829-3, NIMA-829-63, NIMA-829-314, PoP25-063108, PoP25-093112}.
However, because of the large beam and plasma sizes, large when compared to the scale which needs to be resolved \cite{NIMA-909-446}, simulations were limited to reduced models.
These models include the quasi-static approximation \cite{PRST-AB6-061301} or the axially symmetric two-dimensional (2D) geometry.
There is an obvious need of three-dimensional (3D) simulations based on first principles to check the validity of the reduced models and to search for new effects which might have been  missed in these reduced models.
In this paper we report the first fully kinetic 3D simulation of the baseline AWAKE variant.
We show that 2D quasi-static models adequately describe the proton beam dynamics, while the electron injection needs a 3D treatment and is rich in novel effects.

We describe the simulated AWAKE scenario in Sec.\,\ref{s2} and the simulation details in Sec.\,\ref{s3}.
Then we discuss the self-modulation of the proton beam in Sec.\,\ref{s4} and the trapping of externally injected witness electrons in Sec.\,\ref{s5}.
In Sec.\,\ref{s6}, we summarize the main findings.

\section{AWAKE baseline scenario}
\label{s2}

The simulated scenario is close to the scenario used in \cite{NIMA-829-3,NIMA-829-76,PoP25-063108}, except the electron beam has no initial angular spread.
It includes a sharp plasma boundary in the radial direction \cite{NIMA-740-197}, realistic longitudinal density transitions at the plasma cell orifices \cite{JPD51-025203}, a positive density gradient along the plasma cell \cite{NIMA-829-63}, and oblique electron injection at a shallow angle \cite{PoP25-063108}.
The initial angular spread (or emittance) of the electron beam can be neglected and taken as zero, if the spread gained during injection to the wave is much larger than the initial value.
The complete set of physical parameters is listed in Table~\ref{t1}.
Here $m$ is the electron mass, $e>0$ is the elementary charge, $c$ is the speed of light, and $\omega_p$ is the electron plasma frequency.

\begin{table*}[t]
 \caption{ Parameters of the simulated variant. Numbers typed in italic are approximate values.}\label{t1}
 \begin{center}\begin{tabular}{lll}\hline
  Parameter \& notation & Value & Dimensionless value (unit) \\ \hline
  Reference plasma density, $n_0$ & $\textit{7} \times 10^{14}\,\text{cm}^{-3}$ & 1\,($n_0$) \\
  Plasma skin depth, $c/\omega_p$, & 0.2\,mm & 1\,($c/\omega_p$) \\
  Plasma length, $L$ & 10\,m & 50\,000\,($c/\omega_p$) \\
  Transition area half-length, $L_\text{tr}$ & 40\,cm & 2\,000\,($c/\omega_p$) \\
  Plasma radius, $r_p$ & 1.4\,mm & 7\,($c/\omega_p$) \\
  Orifice diameter, $D$ & 10\,mm & 50\,($c/\omega_p$) \\
  Plasma ion-to-electron mass ratio, $M_i$, & 157\,000 & 157\,000\\
  Wavebreaking field, $E_0=mc\omega_p/e$, & \textit{2.54}\,GV/m & 1\,($E_0$) \\
  Uncut driver population, $N_b$ & $3\times 10^{11}$ & $3\times 10^{11}$ \\
  Driver length, $\sigma_{zb}$ & 12\,cm & 600\,($c/\omega_p$) \\
  Driver radius, $\sigma_{rb}$ & 0.2\,mm & 1\,($c/\omega_p$) \\
  Driver energy, $W_b$ & 400\,GeV & \textit{783\,000}\,($mc^2$) \\
  Driver energy spread, $\delta W_b$ & 0.35\,\% & \textit{2740}\,$(mc^2)$ \\
  Driver angular spread, $\delta \alpha_b$, & $4.5 \times 10^{-5}$ & $4.5 \times 10^{-5}$\\
  Maximum driver density, $n_{b0}$ & $\textit{4}\times 10^{12}\,\text{cm}^{-3}$ & \textit{0.0056}\,($n_0$) \\
  Electron beam population, $N_e$ & $1.25\times 10^{9}$ & $1.25\times 10^{9}$ \\
  Electron beam length, $\sigma_{ze}$ & 1.2\,mm & 6\,($c/\omega_p$) \\
  Electron beam radius, $\sigma_{re}$ & 0.25\,mm & 1.25\,($c/\omega_p$) \\
  Electron beam energy, $W_e$ & \textit{16}\,MeV & 32\,$(mc^2)$ \\
  Electron beam energy spread, $\delta W_e$ & 0 & 0 \\
  Electron beam angular spread, $\delta \alpha_e$, & 0 & 0\\
  Maximum electron beam density, $n_{e0}$ & $\textit{1}\times 10^{12}\,\text{cm}^{-3}$ & \textit{0.0015}\,($n_0$) \\
  Injection angle for electron beam, $\alpha_0$, & 0.0028 &  0.0028 \\
  Injection delay, $\xi_e$, & 11.5\,cm & 575\,($c/\omega_p$)  \\
  Intersection of beam trajectories, $z_0$, & \textit{142}\,cm & \textit{7100}\,($c/\omega_p$)  \\
  \hline
 \end{tabular}\end{center}
\end{table*}

We either use cylindrical coordinates $(r, \varphi, z)$ or Cartesian coordinates $(x,y,z)$ with the $z$-axis being the direction of the beam propagation. We additionally consider the co-moving coordinate $\xi = z-ct$.
The longitudinal plasma density profile $n(z)$ is
\begin{equation}\label{e1}
    \begin{array}{ll}
        |z| \le L_\text{tr}: &
            \ds \frac{n_0}{2} \left( 1 + \frac{z/D}{\sqrt{(z/D)^2+0.25}} \right), \\
        L_\text{tr} < z < L - L_\text{tr}: &
            \ds n_0 \left(1 + 0.01 \frac{z-L_\text{tr}}{L - 2 L_\text{tr}} \right), \\
        |z - L| \le L_\text{tr}: &
            \ds \frac{1.01 n_0}{2} \left( 1 - \frac{(z-L)/D}{\sqrt{\left(\frac{z-L}{D}\right)^2+0.25}} \right),
    \end{array}
\end{equation}
and zero otherwise.
The orifice diameter $D$ determines the density profile near the entrance to and exit from the plasma cell (located at $z=0$ and $z=L$).
The plasma is radially uniform within the radius $r_p$ and composed of single ionized rubidium ions.

\begin{figure}[tb]\centering
\centering
\includegraphics{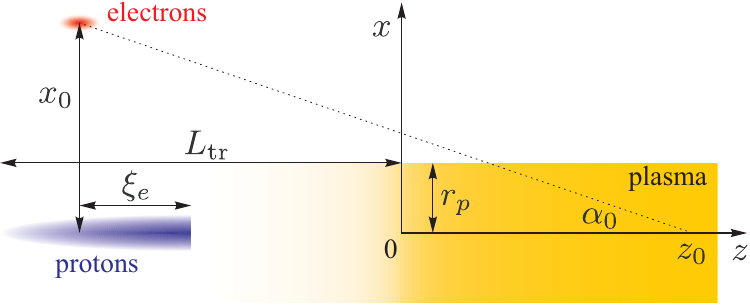}
\caption{Scheme of beam injection into the plasma.}\label{fig1-scheme}
\end{figure}

Before entering the plasma, the beam density is
\begin{eqnarray}\nonumber
 \fl n_b (r, z, t) = 0.5 \, n_{b0} \, e^{-r^2/2 \sigma_{rb}^2} & \left[  1 + \cos \left( \sqrt{\frac{\pi}{2}} \frac{\xi}{\sigma_{zb}}  \right)  \right], \\
\label{e2}
  & -\sigma_{zb} \sqrt{2\pi} < \xi < 0,
\end{eqnarray}
and zero otherwise.
The tail of the driving beam at the simulation start is positioned at $z = -L_\text{tr}$ (Fig.\,\ref{fig1-scheme}).
The cut density profile (\ref{e2}) mimics plasma creation through rapid ionization of a neutral gas by a co-propagating short and intense laser pulse.
The electron beam has the density profile
\begin{eqnarray}\nonumber
    n_e (x,y,z,t) = 0.5 \, n_{e0} \, \exp \left( -\frac{(x-x_0)^2 + y^2}{2 \sigma_{re}^2} \right) \\
\label{e3}
    \times \left[  1 + \cos \left(  \sqrt{\frac{\pi}{2}} \frac{(\xi+\xi_e)}{\sigma_{ze}}  \right)  \right], \quad |\xi+\xi_e| < \sigma_{ze} \sqrt{2\pi},
\end{eqnarray}
and zero otherwise.
Here
\begin{equation}\label{e4}
    x_0 = (z_0 - z) \tan \alpha_0
\end{equation}
is the transverse location of the electron beam center.
All electrons have the same initial momentum $(-\alpha_0 p_0 , 0, p_0)$ with $p_0 = W_e/c$.

\section{Simulations}
\label{s3}
The simulation is realized using the fully relativistic 3D particle-in-cell (PIC) code PSC \cite{JCP318-PSC}, utilizing a standard Boris particle pusher \cite{CFSPPROC-Boris} with second-order quasi-particles, and a finite-difference-time-domain field solver \cite{IEETAP14-FDTD}.
A collection of numerical parameters of the simulation is shown in Table~\ref{t2}.
This full 3D fully kinetic simulation of the AWAKE baseline requires substantial computational resources.
We acquired the considerable amount of 25 Mch at the SuperMUC Phase 1 supercomputer in Munich through a Gauss Center for Supercomputing large scale projects call.
In preparation for the full simulation we perform a multitude of small scale simulations.
These smaller simulations are using the same parameters as the full simulation, except for the spatial resolution which is set to a factor of $1/2$ or $1/4$ of the full resolution of 130 points per plasma wavelength.
This reduces the required number of timesteps by a similar factor.

\begin{table*}[t]
 \caption{ Simulation parameters. Numbers typed in italic are approximate values. }\label{t2}
 \begin{center}\begin{tabular}{lll}\hline
  Parameter \& notation & Value & Dimensionless value (unit)\\ \hline
  Minimum number of grid points per $\lambda_p$ &  & 130\\
  Timestep size & $\textit{1.6} \times 10^{-14}$\,s & \textit{0.025}\,($1/\omega_p$) \\ 
  Spatial grid size, $\Delta x$ & $\textit{9.40}\,\mu$m & \textit{0.047}\,($c/\omega_p$) \\ 
  Spatial grid size, $\Delta y$ & $\textit{9.38}\,\mu$m & \textit{0.047}\,($c/\omega_p$) \\ 
  Spatial grid size, $\Delta z$ & $\textit{9.59}\,\mu$m & \textit{0.048}\,($c/\omega_p$) \\ 
  Initial box size in $x$-direction & 0.8124\,cm & 40.62\,($c/\omega_p$) \\
  Reduced box size in $x$-direction & 0.42\,cm & 21\,($c/\omega_p$) \\
  Box size in $y$-direction & 0.42\,cm & 21\,($c/\omega_p$) \\
  Full box size in $z$-direction & 11.1008\,m & 55504\,($c/\omega_p$) \\
  Galilean window size in $z$-direction & 30.38\,cm & 1519\,($c/\omega_p$) \\
  Initial grid size &   & $864 \times 448 \times 1157120 = \textit{5} \times 10^{11}$\\
  Reduced grid size (Galilean, beams merged) &   & $448 \times 448 \times 32400 = \textit{6} \times 10^9$\\
  Initial number of patches &  & $54 \times 28 \times 72320$\\
  Reduced number of patches (Galilean, beams merged) &  & $28 \times 28 \times 2025$\\
  Number of timesteps &  & 2198134\\
  Quasi-particles per cell for each species &  & 3\\
  Courant-Friedrichs-Lewy number \cite{MA100-CFL} &  & 0.9\\
  Real particles per quasi-particle with weight 1, $N_\text{real}$ &  & $\textit{7.35} \times 10^{7}$\\
  Plasma peak real particles per quasi-particle, $N_\text{real}n_0/n_0$ &  & $\textit{7.35} \times 10^{7}$\\
  Driver peak real particles per quasi-particle, $N_\text{real}n_{b0}/n_0$ &  & $\textit{4.13} \times 10^{5}$\\
  Electron peak real particles per quasi-particle, $N_\text{real}n_{e0}/n_0$ &  & $\textit{1.10} \times 10^{5}$\\
  Average number of quasi-particles in box & & $\textit{7.4} \times 10^9$ \\
  \hline
 \end{tabular}\end{center}
\end{table*}

\begin{figure}[tb]
\centering
\includegraphics[width=194bp]{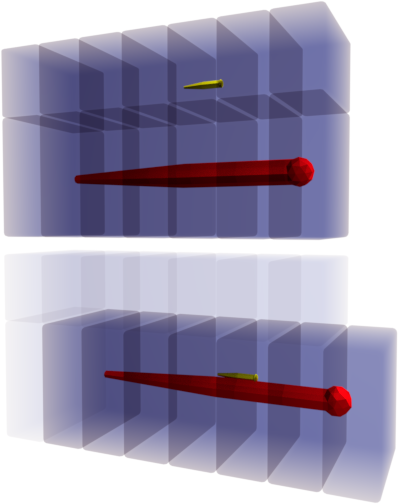}
\caption{The upper figure shows the initial simulation setup containing the ion beam in red and the witness beam in yellow, offset in the $x$-direction. The blue-grey boxes represent the patches. The real simulation uses many more patches in all three directions as given in Table~\ref{t2}. The lower figure shows the configuration of the simulation after the witness and ion beam have merged. The upper patches have been disabled. A patch in front of the beam has been added, while patches in the back of the beam have been disabled. Disabled patches are represented by having a higher transparency.}\label{fig2-patches}
\end{figure}

The PSC offers a powerful mechanism, which provides several key capabilities for this simulation.
This mechanism subdivides the simulation box into several smaller groups of grid points, called patches.
These patches can then be freely distributed over the available resources and can also be activated or deactivated.
Taking full advantage of this dynamic patch system enables three very important ways to save on computational resources.
First, it makes the usage of a moving Galilean window possible.
This window moves at the same speed as the proton beam.
Through this, we reduce the active simulation domain to $1.01 \sqrt{2\pi} \sigma_{zb}\approx 31\,\mathrm{cm}$ in the $z$-direction.
Second, we shrink the simulation domain by $\approx 48\,\%$ after $\approx 17\,\%$ of the simulation by switching off the space occupied by the witness beam after its entry into the plasma cylinder (Fig.\,\ref{fig2-patches}).
Third, we meaningfully reduce load imbalance by continuously rearranging the patches on the participating resources.
Patches with less quasi-particles are grouped and assigned to cores in a way which aims at achieving an equal distribution of quasi-particles to cores.
This is done while maintaining a consecutive spatial order of the patches along a space-filling curve, which is necessary to keep the communication costs low.

The boundary conditions in the lower and upper $z$-direction are fairly straightforward as they do not affect the physics at all.
The reason for this is that no field or quasi-particle can reach the upper boundary as the boundary travels with the same velocity as the particle beam.
The lower boundary is continuously moving as well, which gets rid of any spurious fields and particles before they can return and interact with the simulation.
This makes it reasonable to choose a very simple and predictable reflecting boundary in the positive and negative $z$-direction.
Unfortunately, the transversal direction proves to be very tricky.
After trying out different implementations of universal perfectly-matched layers \cite{IEEETAP44-UPML1,AH-UPML2} and conducting boundaries we come to the conclusion that it is too risky to run the full simulation with these boundaries.
The boundary cells exhibit a certain degree of nonlinear numerical field growth in all tested cases which make the results unreliable.
This may be remedied in future simulations by improvements to the patch mechanism in conjunction with the boundary algorithms.
Spurious currents and the continuously charging wall make it very questionable to just delete exiting particles.
Because of this, we implement periodic boundaries in the transversal direction for particles and fields.
Periodic boundaries are very easy to predict and are very unlikely to behave in an unexpected manner.
Since the plasma provides very strong electromagnetic shielding and the Galilean window guarantees a maximum number of box traversals for electromagnetic waves, there is no effect to the particle dynamics by recurring waves.
Unfortunately, recurring particles are a problem in the case of the electron beam.
Because of the periodic boundary conditions in the transversal direction, witness beam particles can only leave the simulation box by deceleration.
This enables the trapping of witness particles which had already left the plasma after interacting with the beam, particles which, in the experiment, would be lost.
Because of this effect, we can not make proper observations of the captured charge.
An easy solution to this problem would have been to flag each electron quasi-particle which traverses the periodic boundaries.
This would have made it possible to identify and filter these particles during the post-processing step of the simulation.
Unfortunately, this was not implemented for the presented simulation, and the simulation could not be repeated due to its high cost.

The simulation uses three different particle species: plasma electrons, ion beam particles and electron beam particles.
The rubidium ions are assumed to be stationary and are not simulated.
The charge-conserving particle pushing algorithm \cite{CPC135-esirkepov} of the PSC assumes a balanced charge distribution at the start of the simulation.
Not generating any quasi-particles for the rubidium ions therefore acts in the same way as having fixed positive charges at the initial positions of the plasma electrons.
These fixed charges consume no computing resources.
Each kind uses three quasi-particles per cell.
The weight of a quasi-particle is given by $n/n_0$, $n_b/n_0$ and $n_e/n_0$, respectively, where the density is taken at the center of the cell where the quasi-particle is created.
The number of real particles per quasi-particle with a weight of 1 is given by $N_\text{real} = n_0/3 \times \Delta x \times \Delta y \times \Delta z$.
This number is valid for quasi-particles of all species.
It is also the peak number of real particles per plasma quasi-particle, since in this case, $n/n_0=1$, which makes plasma quasi-particles at locations with a peak plasma density have a weight of 1. 
Each cell with a nonzero density value for a specific kind contains three particles of that kind.
We thereby achieve a good resolution of the particle dynamics even in low density areas of the simulation.
These formulas generate quasi-particles with weights which differ by orders of magnitude.
Since we do not use any Monte-Carlo collisions, this does not affect the quality of our results.
On average $\approx 16.3\,\%$ of the quasi-particles are driver particles, $\approx 0.3\,\%$ are electron beam particles and $\approx 83.4\,\%$ are plasma particles.
This makes the number of quasi-particles per cell, which are used to represent the plasma, a dominating factor in determining the cost of the simulation.
We choose to use 3 quasi-particle per cell for the plasma in order to keep the simulation cost reasonable.
Since the plasma exhibits only linear dynamics, this number suffices to accurately represent the physics.
By using 3 quasi-particles for the beam and electron species as well, our simulation has a two orders of magnitude smaller amount of real particles per quasi-particle for these species, and therefore supplies a sufficient resolution for their non-linear dynamics.
The non-peak value of the number of particles per quasi-particle is even lower, which gives the slopes of the Gaussian-distributed beam and electron densities, and therefore the majority of the volume, an even higher resolution.
The plasma quasi-particles have an additional restriction: only cells with a plasma density $> 0.01 n_0$ receive three quasi-particles, other cells receive zero quasi-particles.
This sets a sensible cutoff point for the quasi-particle generation without meaningfully altering the physics.
There is no overlap between the driver and the plasma quasi-particles at the start as the plasma density is below this threshold in the regions where driver particles are setup.

Using the previous parameters, we can calculate the average number of quasi-particles in the box.
We can use this number to give a very simple lower bound for the computational cost of the simulation.
In order to calculate this lower bound we assume that a) a single core can push one million particles per second and that there is no b) communication, c) field solving and d) data output cost.
With these assumptions the computational cost of the simulation can be approximated to be $7.4 \times 10^9 / (10^6/\text{s}) \times 219813\times 1\,\text{c} = 4.5\,\text{Mch}$.
The total computational cost of the final simulation turns out to be about $22\,\text{Mch}$.
Important additional effects, which reduce the overall performance, and increase the cost of the simulation, are periodic check-pointing and restarting times due to the maximum allowed job time on the SuperMUC system as well as unmitigated load imbalance.
From the cost of about 85\,M\,\euro{} for building SuperMUC Phase~1, the total number of cores, which is $18432 \times 8$ (Intel Xeon E5-2680) $+\,\, 820 \times 10$ (Intel Xeon E7-4870) $ = 155656$, and a total runtime of approximately 6 years we can calculate a very conservative price per core-hour of about $85 \times 10^6 / (155656\times 6 \times 356 \times 24) \approx 0.01\,$\euro{}.
This neglects a lot of factors, which increase the cost of each core-hour, like operating costs and system availability.
This lower bound fits well to scientific evaluations \cite{IEEEC42-CHCOST}.
From this follows that the minimal total cost of this simulation amounts to $\approx 220$\,k\euro.
This cost makes multiple simulations at the required resolution prohibitive.

It is extremely valuable to incorporate most of the data analysis into the simulation code itself.
This enables the exploitation of the supercomputer resources for post-processing and reduces the amount of final data, which needs to be saved to disk.
This in turn saves processing time as well, as disk input/output (I/O) is a very time-consuming process.
Every 1000th timestep, the code provides outputs of a variety of smoothed field values with already performed additional processing like averaging and maximum finding.
Furthermore it generates processed particle data like emittance, average locations and respective standard deviations as well as several different kinds of densities for different projections of the configuration and momentum space.
This reduces the output data size immensely and makes it possible to continuously monitor the simulation.
Additionally, ten specific timesteps are saved to disk in their entirety.
These timesteps are chosen to portray important milestones in the evolving simulation, like the orifice entry, the merging of the witness beam, reaching the wakefield maximum, and exiting the plasma through the second orifice.
Another very important data output is the full configuration of 50 particles of each kind every 20 timesteps.
The small number of particles makes it possible to cache this information for prolonged times and perform an output at regular larger-spaced intervals.
Using this data, we are able to perform fine-grained tracking and analysis of the behaviour of the particles.

I/O is generally one of the hardest things to do properly in the context of large scale simulations.
We identify several key parameters of the respective supercomputer and use these parameters for our optimization.
These parameters may differ for different kinds of file systems and may therefore not easily transfer to other supercomputers.
Nonetheless, the general idea may prove helpful for future simulations.
Two very important parameters of this kind are the number of active concurrent file-streams and the number of files per directory for which the I/O performance remains acceptable.
For SuperMUC Phase 1's file system it is necessary to keep both of these numbers below 1000.
It is also of paramount importance to only access each individual directory from a single node at a time.
Using these findings, we carefully designed a parallel output algorithm which serializes the access to directories in an optimal way.
We also tested output libraries, like MPI-IO, but found that due to the PIC data configuration, which consists of large chunks of unscattered data, these libraries did not perform well and did not improve performance meaningfully.
Finally, we are able to achieve a data rate of $105\,\text{GByte}/\text{s}$.
The theoretical machine maximum is given at $125\,\text{GByte}/\text{s}$.
A full checkpoint, which is necessary to be written regularly because of the maximum job time length, had an initial size of $120\,\text{TByte}$ and took $2.5\,\text{h}$ to write.
We reduced the data size to $12\,\text{TByte}$ and are able to write it to disk in three minutes.
This is achieved by exploiting additional features like in-memory compression and optimized data structures. 

The second area of improvement is concerned with the scalability of the code.
Running on 32768 cores simultaneously, will uncover any non-scaling code structure.
We got rid of a slew of unnecessary data duplication and output files.
Through extensive testing, we also identify the optimal amount of cores for our problem.
Due to the need for large amounts of memory as well as an optimized scaling behaviour of the code, we find that 32768 cores is the optimal amount for our case.

\section{Proton beam self-modulation}
\label{s4}
% script to plot in simple_hdf5.py: plot_ezmax_diff_res()
\begin{figure}[tb]
\centering
\includegraphics[]{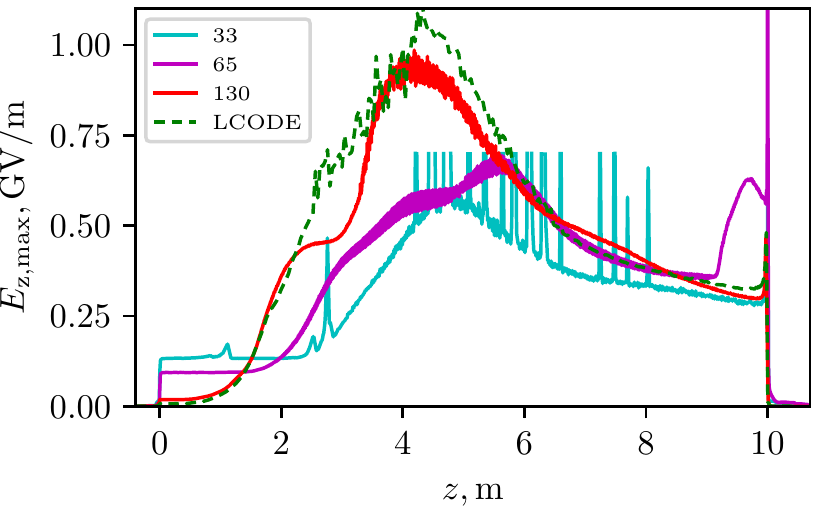}
\caption{Comparison of the maximum accelerating gradient $E_\text{z,max} (z) = \max_{x, y, t}E_\text{z}(x, y, z, t)$ (PSC) or $E_\text{z,max} (z) = \max_{r, t}E_\text{z}(r, z, t)$ (LCODE) for different resolutions using the PSC and the quasi-static code LCODE. The numbers in the legend for the first three lines refer to the number of points per $\lambda_p$ when using the PSC. The dashed line refers to the reference simulation with LCODE using $\approx 628$ number of points per $\lambda_p$. In order to reduce clutter, resulting from a lot of noise, the 33 points per $\lambda_p$ results have been capped at $0.7\,\text{GV/m}$ maximum field value.}\label{fig3-ezmax}
\end{figure}

The plasma transforms the long proton beam into a train of short micro-bunches, which resonantly drive the plasma wave.
A good quantitative measure of this process is the generated longitudinal electric field (Fig.\,\ref{fig3-ezmax}).
From comparing runs with different resolutions, we find that a fine grid is necessary in order to adequately represent the physical processes in the system.
For a sufficiently small grid step size of $\lambda_p/130 \approx 0.05c/\omega_p$, the results of the 3D PSC simulation closely agree with the high resolution reference simulation of the 2D quasi-static code LCODE \cite{PRST-AB6-061301,NIMA-829-350}, with a grid step size of $0.01c/\omega_p$.

% Script to plot in particle_uid.py: plot_colormap_ions_dual()
\begin{figure*}[tb]
\centering
\includegraphics[]{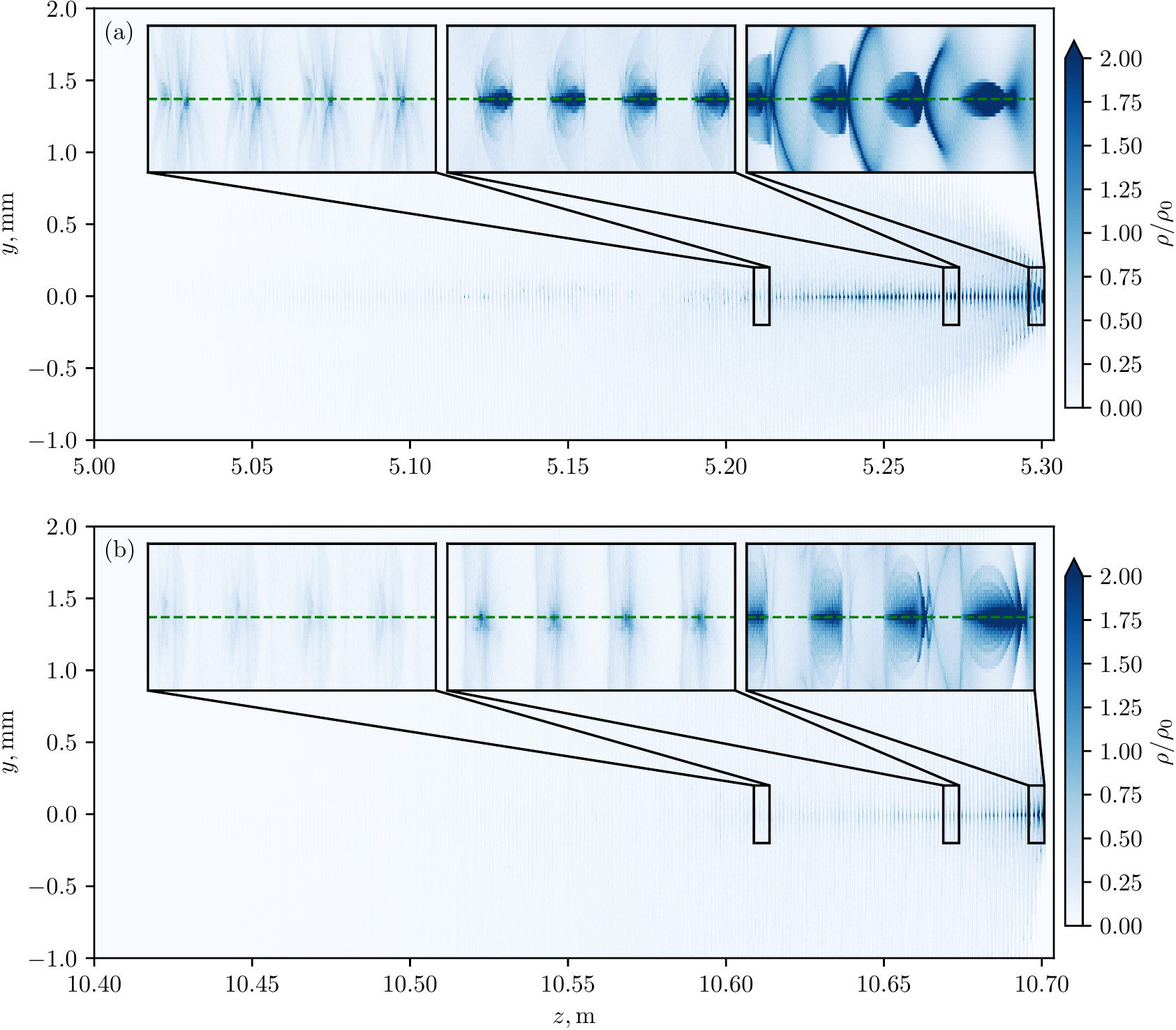}
\caption{Proton density $\rho(y,z)$ integrated in $x$-direction (a) in the middle of and (b) near the exit from the plasma section. The shown figure corresponds to the perspective of the electron beam, which is looking down onto the driver. The colorbar unit $\rho_\text{0}$ corresponds to the density at the center of the unperturbed beam. The green dashed lines in the insets show the propagation axis.}\label{fig4-bunching}
\end{figure*}

Both high-resolution runs show the typical stages of wakefield evolution as the beam propagates through the plasma \cite{PoP22-103110}.
During the first meter in the plasma the wakefield stays at approximately the seed level.
This is then followed by exponential growth, which changes into non-exponential growth after 2\,m.
The beam is fully micro-bunched at about 4\,m [Fig.\,\ref{fig4-bunching}(a)] into the plasma and at this point it excites the strongest wakefield.
After this, the wakefield decays because the micro-bunches do not fully reside in the focusing phase of the wave, and their defocused parts gradually erode \cite{PoP18-024501} [Fig.\,\ref{fig4-bunching}(b)].
The field spike near the exit appears, because the plasma wavelength slightly increases as the plasma density decreases.
The micro-bunches then gradually shift into the stronger decelerating phase of the wave.
Particles in this phase drive the wake more efficiently, which increases the field values.
However, this occurs at the expense of a weaker to non-existing radial focusing, which is why the reduced plasma density can only boost the wakefield for a short distance.

We particularly emphasize that we did not find any sign of beam hosing (Fig.\,\ref{fig4-bunching}).
Even the tail bunches, which are not perfectly symmetric, follow the head bunches perfectly.
Theory \cite{PoP20-056704,PRL112-205001} and simulations of shorter beams \cite{PRL104-255003,PoP19-063105,PRST-AB16-041301,PRL112-205001} predicted that the seeded self-modulation suppresses the hosing, but this is the first observation of no hosing occurring in full scale simulations of AWAKE.
The seeding of the hosing instability in the simulation is even stronger than in reality, because the number of beam quasi-particles in the simulation is smaller than the number of protons in the real beam (Table~\ref{t2}) \cite{PRST-AB16-041301}.
Therefore, the hosing instability has an even lower probability to develop in experiments.

\begin{figure*}[tb]
\centering
\includegraphics{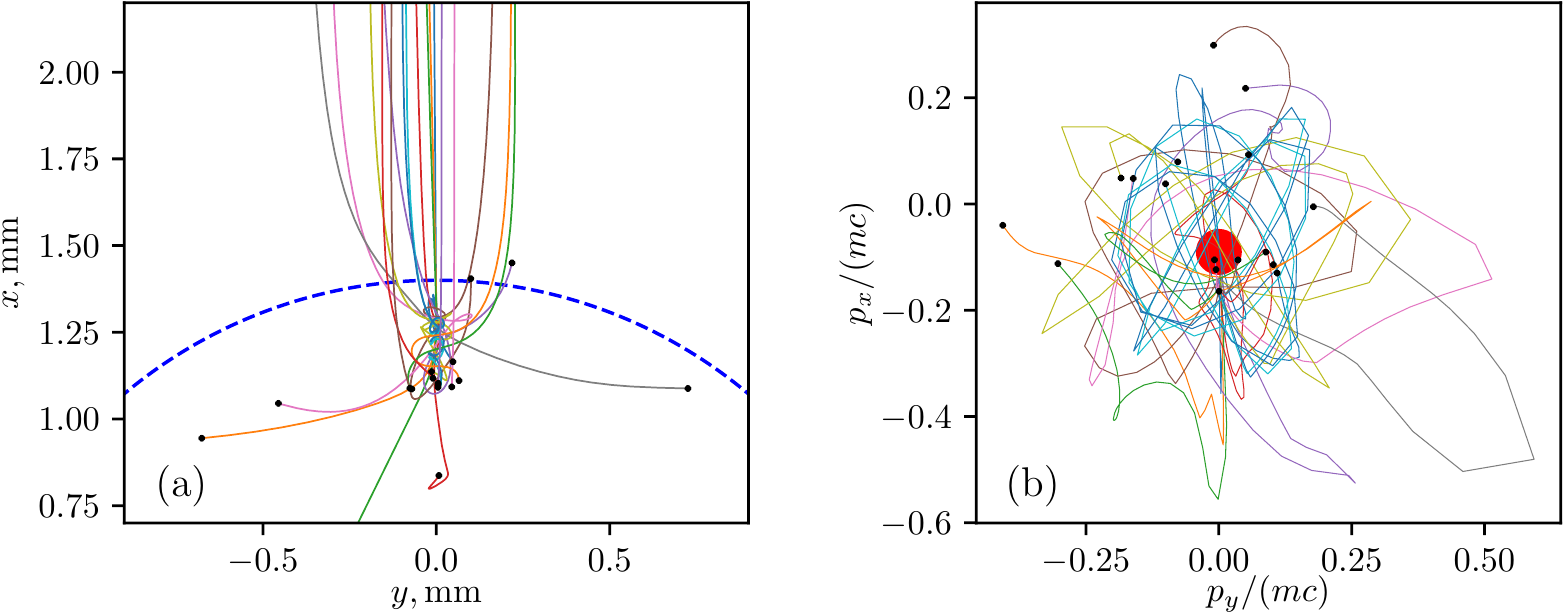}
\caption{Parts of typical electron trajectories in (a) real and (b) momentum space. The small red circle in (b) corresponds to a beam with radius $\sigma_{re}$, energy $W_e$, and normalized emittance 10\,mm\,mrad.}\label{fig5-scattering}
\end{figure*}

\section{Electron injection}
\label{s5}

As the electron bunch approaches and crosses the plasma boundary, it experiences a strong force which is caused by the redistribution of charges in the plasma (Fig.\,\ref{fig5-scattering}).
The force is both focusing the bunch and attracting the bunch to the plasma boundary.
The force action starts in the vacuum and increases as the bunch approaches the boundary [Fig.\,\ref{fig5-scattering}(a)].
In the plasma, the force is strong enough to keep most of the bunch tightly focused.
The transverse momentum, delivered by this force to individual electrons, depends on the electron position in the bunch and is very large, which makes the beam emittance blow up during the boundary crossing [Fig.\,\ref{fig5-scattering}(b)].

\begin{figure*}[tb]
\centering
\includegraphics{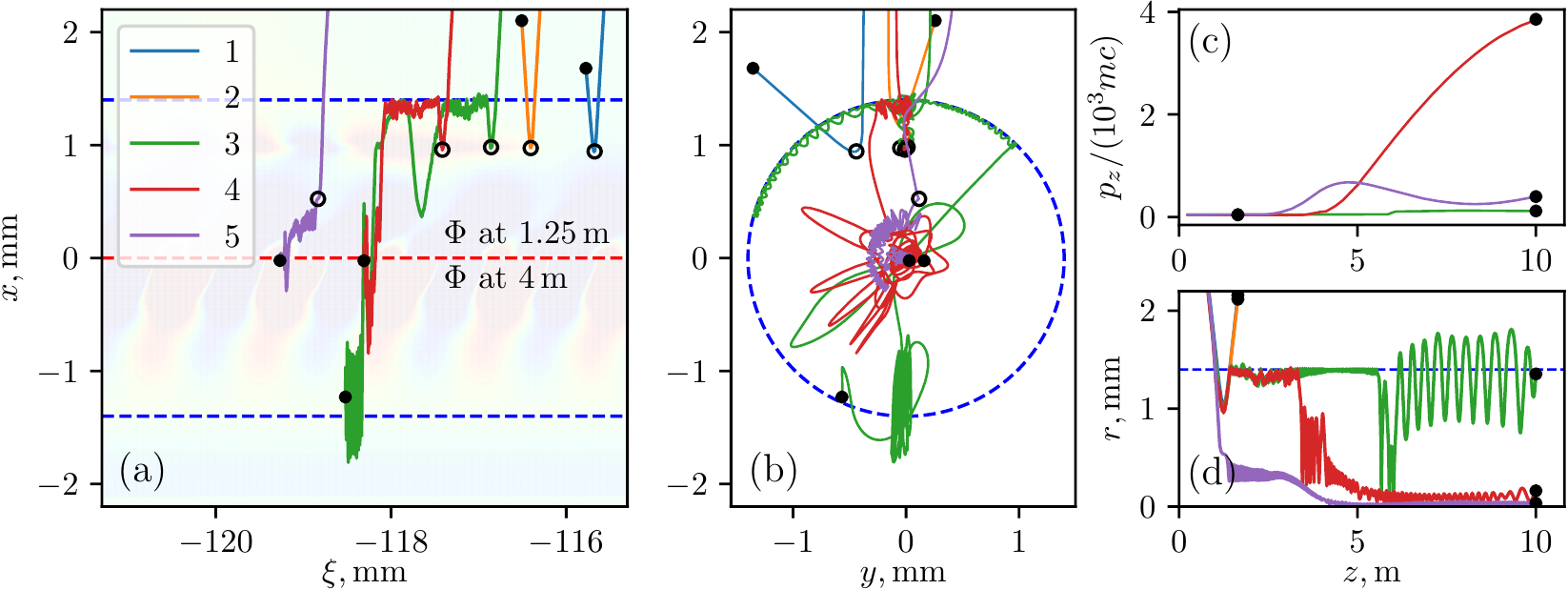}
\caption{Typical electron trajectories in (a), (b) real space, (c) longitudinal momentum $p_z$ and (d) radial position $r$, versus the propagation distance. Black dots show the endpoints of the drawn trajectories (not of the actual particle trajectories), open circles show the particle positions at $z=1.25$\,m. Dashed blue lines show the plasma boundary. Background colors in (a) show the wakefield potential $\Phi (x,\xi)$ at the midplane $y=0$ for $z=1.25$\,m (top) and $z=4$\,m (bottom). The potential $\Phi > 0$ in regions with a red background color (wells for electrons) and $\Phi < 0$ in regions with a blue background color (humps for electrons). Optimal electron acceleration happens in regions with zero-crossings of $\Phi$ from hump to well.
}\label{fig6-trajectories}
\end{figure*}

Most of the electrons in the beam head are simply reflected from the plasma (Fig.\,\ref{fig6-trajectories}).
They enter the plasma, experience a positive radial force, turn around and travel away from the plasma.
An example for this behavior is particle 1 in Fig.\,\ref{fig6-trajectories}.
During their time in the plasma, they create a wakefield that can focus the rest of the beam.
The wakefield force is the gradient of the wakefield potential $\Phi$ shown in Fig.\,\ref{fig6-trajectories}(a).
Here, red regions are potential wells for the electrons, and blue regions are potential humps.
As we see from the potential map, the head particles are reflected from the proton beam wakefield, which is mostly repelling for electrons during the first two meters of propagation.

Particles further downstream (like particles 2--4) are confined by the potential well created by upstream particles and can even make several small-amplitude transverse oscillations in the well.
This potential well traps most of the beam electrons and makes them move in a very similar manner.
The transverse positions of particles 2--4 at $z=1.25$\,m almost coincide [Fig.\,\ref{fig6-trajectories}(b)] in spite of their different initial positions and their acquired transverse momenta.
The potential well follows the transverse position of the electron beam head and eventually comes out of the plasma, pulling the beam body along with it.
Most of the electrons trapped in the well have sufficient transverse momenta and are able to overcome the plasma attraction and fly away (particle 2).
A few, however, remain close to the plasma and oscillate near the boundary (particles~3 and~4).

As the proton beam self-modulates, the potential wells produced by the proton driver become deeper and wider, eventually reaching the plasma boundary and attracting electrons to inner plasma areas (particle 4 at $z \approx 3.5$\,m).
Electrons in opposite wakefield phases (potential humps in the inner area) get locked between the hump and the boundary and oscillate there until they drift to the well.
This eventual drifting is possible because of the difference between the wakefield phase velocity and the electron longitudinal velocity (particle 3, which enters the central area only at $z \approx 5.5$\,m).

Shortly after the self-modulation, the driver wakefield has the following structure: The wave phase in the region with $r \gtrsim c/\omega_p$, which is driven by defocused protons, is opposite to the wave phase in the region around $r=0$, which is driven by the remaining protons.
Because of this structure, electrons can oscillate for a time around a non-zero radial position inside the plasma (particle 4 at $z \approx 4$\,m), before they are finally able to reach the near-axis region (particle~4), return back to the boundary (particle 3), or get lost (not shown).
Close to the axis, the wakefield is strong, and the electron energy quickly grows as high as 2\,GeV (particle~4).
There are, however, only a few electrons exhibiting a behaviour like this.

Since all wakefields, including the wakefield the electron bunch generates in its rear part, are oscillating with the same plasma wavelength, there is a chance for electrons to squeeze through the boundary wakefields and to directly arrive at the axis (particle~5).
These particles are very small in number, as well.

\section{Summary}
\label{s6}

Even at the minimum acceptable resolution, full 3D simulations of proton-driven plasma wakefield acceleration requires huge computational resources and should give way to reduced models wherever possible.
The seeded self-modulation of proton beams is one such effect, which can reliably be simulated with axially symmetric quasi-static codes.
In contrast, however, electron injection into the wakefield needs 3D simulations even if the electron bunch charge is small compared to the proton beam charge.
When a low-energy electron bunch crosses the plasma boundary, it induces its own wakefield in the plasma, which, in combination with peripheral proton wakefield, results in the reflection of most of the beam from the plasma and blowing up the angular spread of the remaining part.
This effect may be responsible for the lower amount of accelerated charge \cite{Nat.561-363}, compared to the predictions made with reduced models \cite{NIMA-829-3}, in experiments.

\ack

This work was supported by the Cluster-of-Excellence Munich Centre for Advanced Photonics (MAP) and the Gauss Centre for Supercomputing (GCS), project PLASMA SIMULATION CODE, LRZ-ID pr84me, and by the Russian Fund for Basic Research, project 19-02-00243. N. M., K. B., F. D. and H. R. acknowledge the hospitality of the Arnold Sommerfeld Center for Theoretical Physics at the Ludwig Maximilians University.


\section*{References}
\begin{thebibliography}{88}
 \bibitem{PoP18-103101}
	A. Caldwell, K. V. Lotov,
	Plasma wakefield acceleration with a modulated proton bunch.
	Phys. Plasmas 18, 103101 (2011).
\bibitem{RAST9-85}
    E. Adli and P. Muggli,
    Proton-Beam-Driven Plasma Acceleration.
    Reviews of Accelerator Science and Technology 9, 85 (2016).
\bibitem{EPJC76-463}
    A.Caldwell and M.Wing,
    VHEeP: a very high energy electron-proton collider.
    Eur. Phys. J. C \textbf{76}, 463 (2016).
\bibitem{NIMA-829-3}
    A. Caldwell, E. Adli, L. Amorim, R. Apsimon, T. Argyropoulos, R. Assmann, A.-M. Bachmann, F. Batsch, J. Bauche, V.K. Berglyd Olsen, et al.,
	Path to AWAKE: Evolution of the concept.
	Nuclear Instr. Methods A 829, 3 (2016).
\bibitem{NIMA-829-76}
    E. Gschwendtner, E. Adli, L. Amorim, R. Apsimon, R. Assmann, A.-M. Bachmann, F. Batsch, J. Bauche, V.K. Berglyd Olsen, M. Bernardini, et al.,
	AWAKE, The Advanced Proton Driven Plasma Wakefield Acceleration Experiment at CERN.
	Nuclear Instr. Methods A 829, 76 (2016).
\bibitem{PPCF60-014046}
	P.Muggli, E.Adli, R.Apsimon, F.Asmus, R.Baartman, A-M.Bachmann, M.Barros Marin, F.Batsch, J.Bauche, V.K.Berglyd Olsen, et al.,
	AWAKE readiness for the study of the seeded self-modulation of a 400 GeV proton bunch.
	Plasma Phys. Control. Fusion 60, 014046 (2018).
\bibitem{PRL122-054801}
    M. Turner, E. Adli, A. Ahuja, O. Apsimon, R. Apsimon, A.-M. Bachmann, M. Barros Marin, D. Barrientos, F. Batsch, J. Batkiewicz, et al.,
    Experimental Observation of Plasma Wakefield Growth Driven by the Seeded Self-Modulation of a Proton Bunch.
    Phys. Rev. Lett. 122, 054801 (2019).
\bibitem{PRL122-054802}
    E. Adli, A. Ahuja, O. Apsimon, R. Apsimon, A.-M. Bachmann, D. Barrientos, M. M. Barros, J. Batkiewicz, F. Batsch, J. Bauche, et al.,
    Experimental Observation of Proton Bunch Modulation in a Plasma at Varying Plasma Densities.
    Phys. Rev. Lett. 122, 054802 (2019).
\bibitem{Nat.561-363}
    E. Adli, A. Ahuja, O. Apsimon, R. Apsimon, A.-M. Bachmann, D. Barrientos, F. Batsch, J. Bauche, V.K. Berglyd Olsen, M. Bernardini et al.,
    Acceleration of electrons in the plasma wakefield of a proton bunch.
    Nature 561, 363 (2018).

\bibitem{JPP78-347}
	G. Xia, R. Assmann, R. A. Fonseca, C. Huang, W. Mori, L. O. Silva, J. Vieira, F. Zimmermann, and P. Muggli,
	A proposed demonstration of an experiment of proton-driven plasma wakefield acceleration based on CERN SPS,
	J. Plasma Phys. 78(4), 347 (2012).
\bibitem{PRL109-145005}
    J.Vieira, R.A.Fonseca, W.B.Mori, and L.O.Silva,
    Ion Motion in Self-Modulated Plasma Wakefield Accelerators.
    Phys. Rev. Lett. \textbf{109}, 145005 (2012).
\bibitem{PoP20-013102}
	K.V.Lotov, A.Pukhov, and A.Caldwell,
	Effect of plasma inhomogeneity on plasma wakefield acceleration driven by long bunches.
	Phys. Plasmas 20(1), 013102 (2013).
\bibitem{PRST-AB16-041301}
	K.V.Lotov, G.Z.Lotova, V.I.Lotov, A.Upadhyay, T.Tuckmantel, A.Pukhov, A.Caldwell,
	Natural noise and external wakefield seeding in a proton-driven plasma accelerator.
	Phys. Rev. ST Accel. Beams \textbf{16}, 041301 (2013).
\bibitem{PoP20-083119}
    K.V.Lotov,
    Excitation of two-dimensional plasma wakefields by trains of equidistant particle bunches.
    Phys. Plasmas 20, 083119 (2013).
\bibitem{PoP20-103111}
    C.Siemon, V.Khudik, S.A.Yi, A.Pukhov, and G.Shvets,
    Laser-seeded modulation instability in a proton driver plasma wakefield accelerator.
    Phys. Plasmas \textbf{20}, 103111 (2013).
\bibitem{PoP21-056705}
    J. Vieira, R. A. Fonseca, W. B. Mori, and L. O. Silva,
    Ion motion in the wake driven by long particle bunches in plasmas.
    Phys. Plasmas \textbf{21}, 056705 (2014).
\bibitem{PRL112-194801}
	K.V.Lotov, A.P.Sosedkin, A.V.Petrenko,
	Long-Term Evolution of Broken Wakefields in Finite-Radius Plasmas.
	Phys. Rev. Lett. 112, 194801 (2014).
\bibitem{PPCF56-084013}
	R.Assmann, R.Bingham, T.Bohl, C.Bracco, B.Buttenschon, A.Butterworth, A.Caldwell, S.Chattopadhyay, S.Cipiccia, E.Feldbaumer, et al.,
	Proton-driven plasma wakefield acceleration: a path to the future of high-energy particle physics.
	Plasma Phys. Control. Fusion \textbf{56}, 084013 (2014).
\bibitem{PoP21-083107}
	K.V.Lotov, V.A.Minakov, and A.P.Sosedkin,
	Parameter sensitivity of plasma wakefields driven by self-modulating proton beams.
	Phys. Plasmas 21, 083107 (2014).
\bibitem{PoP21-123116}
	K.V.Lotov, A.P.Sosedkin, A.V.Petrenko, L.D.Amorim, J.Vieira, R.A.Fonseca, L.O.Silva, E.Gschwendtner, and P.Muggli,
	Electron trapping and acceleration by the plasma wakefield of a self-modulating proton beam.
	Phys. Plasmas 21, 123116 (2014).
\bibitem{NIMA-829-63}
	A. Petrenko, K. Lotov, A. Sosedkin,
	Numerical Studies of Electron Acceleration Behind Self-Modulating Proton Beam in Plasma with a Density Gradient.
	Nuclear Instr. Methods A 829, 63 (2016).
\bibitem{NIMA-829-314}
	M. Turner, A. Petrenko, B. Biskup, S. Burger, E. Gschwendtner, K.V. Lotov, S. Mazzoni, H. Vincke,
	Indirect Self-Modulation Instability Measurement Concept for the AWAKE Proton Beam.
	Nuclear Instr. Methods A 829, 314 (2016).
\bibitem{PRAB21-011301}
    V.K. Berglyd Olsen, E. Adli, and P. Muggli,
    Emittance preservation of an electron beam in a loaded quasilinear plasma wakefield.
    Phys. Rev. Accel. Beams 21, 011301 (2018).
\bibitem{PoP25-063108}
    A.A.Gorn, P.V.Tuev, A.V.Petrenko, A.P.Sosedkin, and K.V.Lotov,
    Response of narrow cylindrical plasmas to dense charged particle beams.
    Phys. Plasmas 25, 063108 (2018).
\bibitem{PoP25-093112}
    V.A. Minakov, M. Tacu, A.P. Sosedkin, and K.V. Lotov,
    Witness emittance growth caused by driver density fluctuations in plasma wakefield accelerators.
    Phys. Plasmas 25, 093112 (2018).
\bibitem{NIMA-909-446}
    K.V. Lotov,
    AWAKE-related benchmarking tests for simulation codes.
    Nuclear Instr. Methods A 909, 446 (2018).
\bibitem{PRST-AB6-061301}
    K.V.Lotov,
    Fine wakefield structure in the blowout regime of plasma wakefield accelerators.
    Phys. Rev. ST - Accel. Beams \textbf{6}, 061301 (2003).
\bibitem{NIMA-740-197}
    E.Oz, P.Muggli,
    A novel Rb vapor plasma source for plasma wakefield accelerators.
    Nucl. Instr. Meth. A \textbf{740}, 197 (2014).
\bibitem{JPD51-025203}
    G. Plyushchev, R. Kersevan, A. Petrenko, and P. Muggli,
    A rubidium vapor source for a plasma source for AWAKE.
    J. Phys. D: Appl. Phys. 51, 025203 (2018).
\bibitem{JCP318-PSC}
    K. Germaschewski, W. Fox, S. Abbott, N. Ahmadi, K. Maynard, L. Wang, H. Ruhl, A. Bhattacharjee,
    The Plasma Simulation Code: A modern particle-in-cell code with patch-based load-balancing.
    Journal of Computational Physics \textbf{318}, 305-326 (2016).
\bibitem{CFSPPROC-Boris}
    J.P.Boris,
    Relativistic plasma simulation-optimization of a hybrid code.
    Proceedings of the 4th Conference on Numerical Simulation of Plasmas. Naval Res. Lab., Washington, D.C., 367 (1970).
\bibitem{IEETAP14-FDTD}
    K. Yee,
    Numerical solution of initial boundary value problems involving maxwells equations in isotropic media.
    IEEE Transactions on Antennas and Propagation \textbf{14(3)}, 302307 (1966).
\bibitem{MA100-CFL}
    R. Courant, K. Friedrichs, H. Lewy,
    Ueber die partiellen Differenzengleichungen der mathematischen Physik.
    Mathematische Annalen, \textbf{100} (1), 32–74 (1928).
\bibitem{IEEETAP44-UPML1}
    S. D. Gedney,
    An anisotropic perfectly matched layer-absorbing medium for the truncation of FDTD lattices.
    IEEE Transactions on Antennas and Propagation, \textbf{44}, 1630-1639 (1996).
\bibitem{AH-UPML2}
    A. Taflove, S. C. Hagness,
    Computational Electrodynamics - The Finite Difference Time Domain Method (2nd ed.).
    Artech House (2000).
\bibitem{CPC135-esirkepov}
    T.Zh. Esirkepov,
    Exact charge conservation scheme for Particle-in-Cell simulation with an arbitrary form-factor.
    Computer Physics Communications \textbf{135}, 144-153 (2001).
\bibitem{IEEEC42-CHCOST}
    E. Walker,
    The Real Cost of a {CPU} Hour.
    {IEEE} Computer \textbf{42 (4)}, 35--41 (2009).
\bibitem{NIMA-829-350}
	A.P.Sosedkin, K.V.Lotov,
	LCODE: A parallel quasistatic code for computationally heavy problems of plasma wakefield acceleration.
	Nuclear Instr. Methods A \textbf{829}, 350 (2016).
\bibitem{PoP22-103110}
	K. V. Lotov,
	Physics of beam self-modulation in plasma wakefield accelerators.
	Phys. Plasmas \textbf{22}, 103110 (2015).
\bibitem{PoP18-024501}
	K.V.Lotov,
	Controlled self-modulation of high energy beams in a plasma.
	Phys. Plasmas 18(2) (2011) 024501.
\bibitem{PoP20-056704}
    C.B.Schroeder, C.Benedetti, E.Esarey, F.J.Gruner, and W.P.Leemans,
    Coherent seeding of self-modulated plasma wakefield accelerators.
    Phys. Plasmas \textbf{20}, 056704 (2013).
\bibitem{PRL112-205001}
    J. Vieira, W. B. Mori, and P. Muggli,
    Hosing Instability Suppression in Self-Modulated Plasma Wakefields.
    Phys. Rev. Lett. \textbf{112}, 205001 (2014).
\bibitem{PRL104-255003}
	N.Kumar, A.Pukhov, and K.Lotov,
	Self-modulation instability of a long proton bunch in plasmas.
	Phys. Rev. Lett. \textbf{104}, 255003 (2010).
\bibitem{PoP19-063105}
    J. Vieira, Y. Fang, W. B. Mori, L. O. Silva, and P. Muggli,
    Transverse self-modulation of ultra-relativistic lepton beams in the plasma wakefield accelerator.
    Phys. Plasmas \textbf{19}, 063105 (2012).
\end{thebibliography}
\end{document}